\pdfoutput=1
\documentclass[twoside,fleqn,12pt,headings=normal]{scrartcl}

\usepackage{scrlayer-scrpage}
\ofoot{}
\newcommand{\shorttitle}{\cohead}
\newcommand{\shortauthor}{\cehead}
\newcommand{\affiliation}{\date}
\newcommand{\aff}{\textsuperscript}
\newcommand{\corresp}[1]{\thanks{Email address for correspondence: #1}\hspace{0.5ex}}
\newcommand{\email}[1]{\href{mailto:#1}{\texttt{#1}}}
\renewenvironment{abstract}{\noindent}{}
\usepackage{natbib}
\addtokomafont{title}{\LARGE}
\addtokomafont{author}{\large}
\setkomafont{date}{\small}
\setkomafont{publishers}{\small}
\setcapindent{1em}
\setkomafont{captionlabel}{\small\sffamily\bfseries}
\addtokomafont{caption}{\small}
\RedeclareSectionCommand[%
beforeskip=4pt plus 2pt minus 1pt,%
afterskip=4pt plus 2pt minus 1pt%
]{section}
\RedeclareSectionCommand[%
beforeskip=4pt plus 2pt minus 1pt,%
afterskip=2pt plus 1pt minus 1pt%
]{subsection}
\makeatletter
\renewcommand*{\@maketitle}{%
  \global\@topnum=\z@
  \setparsizes{\z@}{\z@}{\z@\@plus 1fil}\par@updaterelative
  \ifx\@titlehead\@empty \else
    \begin{minipage}[t]{\textwidth}
      \usekomafont{titlehead}{\@titlehead\par}%
    \end{minipage}\par
  \fi
  \null
  \vskip -4em\strut
  \begin{center}%
    \ifx\@subject\@empty \else
      {\usekomafont{subject}{\@subject \par}}%
      \vskip 1.5em
    \fi
    {\usekomafont{title}{\@title \par}}
    \vskip .5em
    {\ifx\@subtitle\@empty\else\usekomafont{subtitle}\@subtitle\par\fi}%
    \vskip 1em
    {%
      \usekomafont{author}{%
        \lineskip .5em%
        \begin{tabular}[t]{c}
          \@author
        \end{tabular}\par
      }%
    }%
    \vskip 1em%
    {\usekomafont{date}{\@date \par}}%
    \vskip \z@ \@plus 1em
    {\usekomafont{publishers}{\@publishers \par}}%
    \ifx\@dedication\@empty \else
      \vskip 2em
      {\usekomafont{dedication}{\@dedication \par}}%
    \fi
  \end{center}%
  \par
  \vskip 0.5em
}%
\makeatother
\publishers{(Preprint of a  Journal of Fluid Mechanics article, \href{http://dx.doi.org/10.1017/jfm.2019.631}{\texttt{doi:10.1017/jfm.2019.631})}}

\usepackage[T1]{fontenc}
\usepackage{amsmath,amssymb}
\usepackage{graphicx}
\graphicspath{{figures/},{./}}
\usepackage[range-phrase = --,retain-unity-mantissa = false,exponent-product = \cdot,separate-uncertainty]{siunitx}

\usepackage{hyperref}

\newcommand{\diff}{\mathrm{d}}
\renewcommand{\vec}{\mathbf}

\hypersetup{%
  pdfauthor={Adrian Daerr, Juliette Courson, Margaux Abello, Wladimir Toutain, Bruno Andreotti},
  pdftitle={The charmed string: self-supporting loops through air drag},
  pdfsubject={Fluid mechanics,Fluid-structure interactions},
  pdfkeywords={Drag,turbulent boundary layer,incompressible flow,supersonic-subsonic transition},
  colorlinks=false,pdfborder={0 0 0},%
  linkcolor=black,citecolor=black,%
  filecolor=blue,urlcolor=blue,%
}

\shorttitle{The charmed string: self-supporting loops through air drag}

\shortauthor{A. Daerr, J. Courson, M. Abello, W. Toutain, B. Andreotti}

\title{The charmed string: self-supporting loops through air drag}

\author{%
Adrian Daerr\aff{1}%
\corresp{\email{adrian.daerr@univ-paris-diderot.fr}},
Juliette Courson\aff{1},
Margaux Abello\aff{1},
Wladimir Toutain\aff{1},
\and Bruno Andreotti\aff{2}%
}

\affiliation{%
  \aff{1}Mati{\`e}re et Syst{\`e}mes Complexes, UMR 7057
  Universit{\'e} de Paris -- CNRS, 10 rue Alice Domon et L{\'e}onie
  Duquet, 75013 Paris, France.
\vskip0pt
  \aff{2}Laboratoire de Physique de l'ENS, UMR 8550 Ecole Normale
  Sup{\'e}rieure -- CNRS -- Universit{\'e}~de~Paris --
  Sorbonne~Universit{\'e}, 24 rue Lhomond, 75005 Paris, France.%
}

\begin{document}

\maketitle

\begin{abstract}
  The string shooter experiment uses counter-rotating pulleys to
  propel a closed string forward. Its steady state exhibits a
  transition from a gravity dominated regime at low velocity towards a
  high velocity regime where the string takes the form of a
  self-supporting loop. Here we show that this loop of light string is
  not suspended in the air due to inertia, but through the
  hydrodynamic drag exerted by the surrounding fluid, namely air. We
  investigate this drag experimentally and theoretically for a smooth
  long cylinder moving along its axis. We then derive the equations
  describing the shape of the string loop in the limit of vanishing
  string radius. The solutions present a critical point, analogous to
  a hydraulic jump, separating a supercritical zone where the wave
  velocity is smaller than the rope velocity, from a subcritical zone
  where waves propagate faster than the rope velocity. This property
  could be leveraged to create a white hole analogue similar to what
  has been demonstrated using surface waves on a flowing fluid. Loop
  solutions that are regular at the critical point are derived,
  discussed and compared to the experiment. In the general case,
  however, the critical point turns out to be the locus of a sharp
  turn of the string, which is modelled theoretically as a
  discontinuity. The hydrodynamic regularisation of this geometrical
  singularity, which involves non-local and added mass effects, is
  discussed based on dimensional analysis.
\end{abstract}


\section{Introduction}
In the broad field of fluid structure interactions, the study of
flexible slender objects in axial flow has been motivated by many
different engineering problems from nuclear reactors to aeronautics
\citep{paidoussis2016}, and also by its importance in living
organisms, for example for the propulsion of swimmers
\citep{gazzola2015}. In the textile industry, ring spinning, air-jet
weft insertion and demand for high weaving speed have triggered
research on the dynamics of light fibres. Here we show that the
hydrodynamic drag of air is a crucial ingredient in the dynamics of
the string shooter, a physics toy in which a closed loop of flexible
string is longitudinally entrained. This toy was popularised online by
\citet{yeany2014}, and its understanding was one of the problems of
2019's International Physicists' Tournament \citep{ipt}.

We also show that string dynamics formally corresponds to 1D
hydrodynamics, with pressure in the Navier-Stokes equations replaced
by string tension. As opposed to the analogous incompressible flow
through a flexible pipe \citep{doare2002,paidoussis2016} however, the
geometrical inversion --- the solid boundary is moving in a
surrounding liquid (air) at rest, as opposed to fluid flowing inside a
guiding tube --- leads to very different dynamics of the string
shooter because momentum is transferred to the environment. This
motivates a particular focus on the hydrodynamic drag experienced by
the string.

%
%
\begin{figure}
  \includegraphics[width=\textwidth]{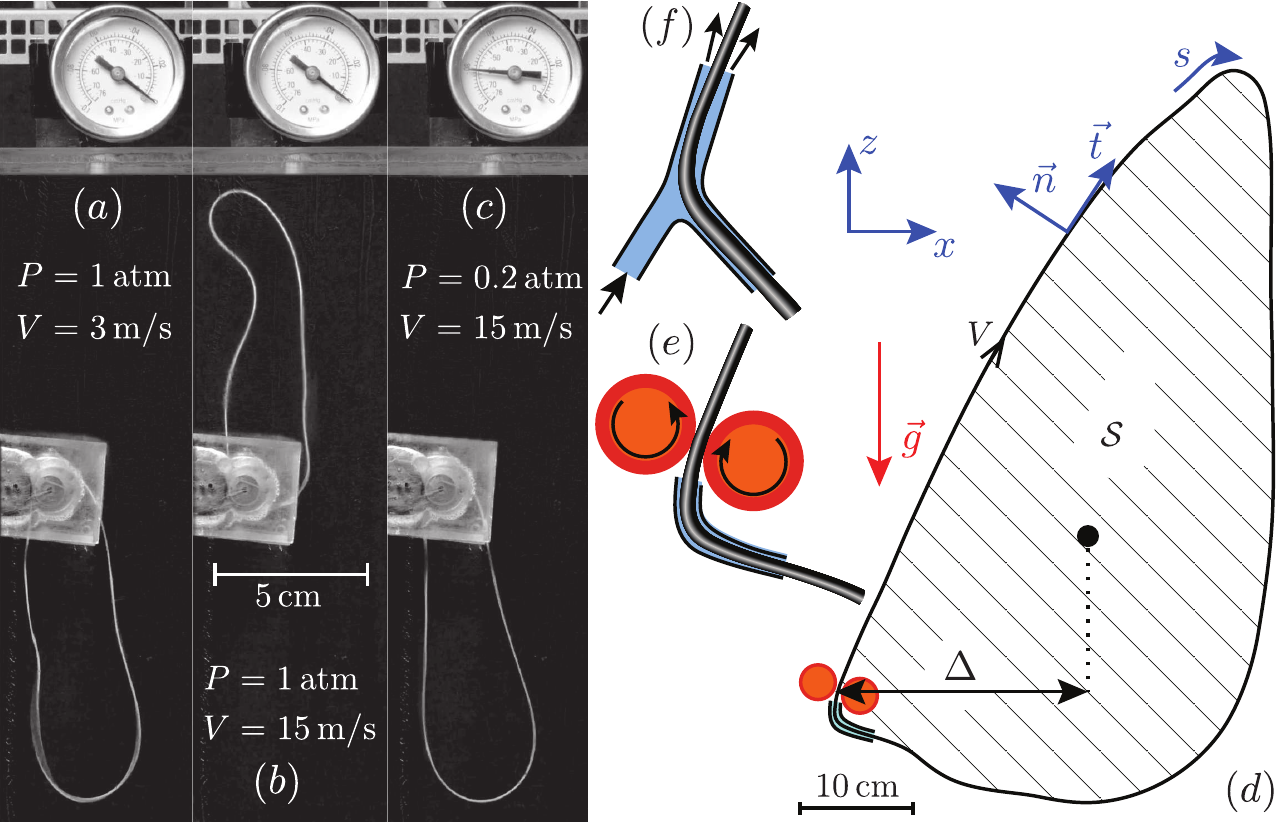}
  \caption{A string loop entrained by two counter-rotating wheels at
    speed $V$ exhibits a transition from (a) a gravity dominated
    regime at low velocity towards (b) a high velocity regime where
    the string takes the form of a self-supporting loop. (c) Reducing
    the pressure from atmospheric pressure to \SI{20}{\kilo\pascal}
    while keeping the high velocity reverts the string back to its
    lowered configuration. Images (a-c) were processed for readability
    by subtracting the background and enhancing contrast. Each image
    is \SI{6.25}{\centi\metre} large. (d) Notations used in the text.
    The string is entrained by either (e) counter-rotating wheels or
    (f) by blowing air along a small portion of it.}
  \vspace{-4 mm}
  \label{fig:VelocityPressure}
\end{figure}

\section{Preliminary observations}
\subsection{Experimental set-up}
A string loop is entrained by rubber wheels mounted on two identical
DC motors rotating at the same velocity but in opposite directions
(figure~\ref{fig:VelocityPressure}). The wheels impose the linear
velocity $V$ and the angle at which the string is ejected. The
transition from a pendent to a self-supporting shape is observed only
for strings of low linear mass. We have used a three-ply cotton
string, with a very low effective elasticity, of radius
$R=\SI{5.5e-4}{\metre}$ and of density
$\rho_s = \SI{300}{\kilo\gram\per\cubic\metre}$. Introducing its
cross-sectional area $A_s =\pi R^2$, the linear mass density is
$ \rho_s A_s=\SI{2.88e-4}{\kilo\gram\per\metre}$ as directly measured
using chemistry scales with a resolution better than one percent. The
length $\ell$ of the string, the ejection angle of the string at the
shooter and the velocity $V$ are our three control parameters.

\subsection{Evidence for the hydrodynamic lift of the string}
Figure~\ref{fig:VelocityPressure} shows low and high velocity steady
state shapes of the string, which are directly related to the fact
that it makes a loop. Indeed, a very long, unclosed string entrained
by the same two wheels traces a long inverted catenary trajectory, as
expected~\citep{biggins2014}. The role of air becomes clear when
performing the experiment at reduced pressure
(figure~\ref{fig:VelocityPressure}c): despite high kinetic energy that
would propel a free string many meters up and away (here
$V^2/g \simeq \SI{23}{\metre}$), the looped string hangs down as for
low velocities at atmospheric pressure. Further evidence for the
importance of air drag is the possibility of replacing the wheel
entrainment (figure~\ref{fig:VelocityPressure}e) by an aerodynamic
entrainment (figure~\ref{fig:VelocityPressure}f).

The string can be safely considered as inextensible: in our
experiments the maximum elongation is of order
$ \rho_s A_s V^2/EA_s = \num{1e-3}$, based on the string's measured
stiffness $EA_s = \SI{35}{\newton}$, and tensions of order
$ \rho_s A_s V^2 = \SI{0.03}{\newton}$ for
$V = \SI{10}{\meter\per\second}$. In the steady state, the continuity
equation applied to any piece of inextensible string implies that the
velocity is tangent everywhere to the string and is constant in
modulus. As a consequence, it can be written as $V \vec t$, where $V$
is imposed by the motor. Consider first a simple model in which one
would neglect the hydrodynamic drag exerted by the air on the string.
Then, the string would be a conservative system whose energy is
minimal at equilibrium. The kinetic energy,
$\frac 12\rho_s A_s \ell V^2$, does not depend on the shape of the
string. As a consequence, such a model predicts that the string shape
should always be the one minimising the potential energy, leading to a
solution independent of the driving velocity
\citep[p.~66]{walton1854} --- in contrast to the observations reported
in figure~\ref{fig:VelocityPressure}.

\section{A minimal string shooter model}
\subsection{A minimal model}
The minimal model for the self-sustained string must therefore involve
inertia, gravity, string tension and the drag by the surrounding
fluid. From dimensional analysis, the drag force per unit length
exerted on a string moving along its tangent takes the form:
$-\alpha \rho_f R |\vec v| \vec v$, where $\alpha$ is a dimensionless
`skin' friction, $\rho_f$ the fluid density and $\vec v$ the material
velocity. Assuming inextensibility, the equation of motion reads:
\begin{equation}
\rho_s A_s \frac{\diff \vec v}{\diff t} =\rho_s A_s \vec g + \frac{\partial (T \vec t)}{\partial s} - \alpha \rho_f R |\vec v| \vec v
\end{equation}
where $s$ is the curvilinear coordinate and $\vec t$ the unit tangent
vector (figure~\ref{fig:VelocityPressure}d). Importantly, the
acceleration is the total time derivative of the velocity:
$\frac{\diff}{\diff t} = \frac{\partial}{\partial t} + \vec v \cdot
\vec t \frac{\partial}{\partial s}$. In the steady state, the
continuity equation implies $\vec v = V \vec t$, resulting in the
following force/inertia balance:
\begin{equation}
\rho_s A_s \vec g + \frac{\partial }{\partial s} \left[\left(T-\rho_s A_s V^2 \right) \vec t\,\,\right]- \alpha \rho_f R V^2 \vec t = 0
 \label{NS}
\end{equation}
The effect of inertia is to shift the tension to negative values by a
constant $\rho_s A_s V^2$. We will call $T-\rho_s A_s V^2$ the
effective tension which, contrary to the true tension $T$, can be
negative. We introduce the dimensionless number comparing tension to
inertial forces as:
\begin{equation}
  \mathcal{T}=\frac{T}{  \rho_s A_s V^2}
\end{equation}
As $\sqrt{T/ \rho_s A_s}$ is the velocity of transverse waves, the
Mach number is simply related to the rescaled tension as
$\mathcal{T}^{-1/2}$. The `Mach 1' critical point therefore coincides
with $\mathcal{T}=1$ and separates a supersonic regime at
$\mathcal{T}<1$ from a subsonic regime at $\mathcal{T}>1$. The
transsonic regime $\mathcal{T}\sim 1$ is discussed at the end of this
article.

\subsection{How can the hydrodynamic drag lift the string?}
\label{sec:torque}
Integrating equation~\eqref{NS} over the closed loop, one sees that the
weight $\rho_s A_s \ell \vec g$ must be balanced by the discontinuity
of the effective tension $(T-\rho_s A_s V^2)\vec t$ across the driving
point. Indeed, the hydrodynamic drag is constant and tangent to the
string and has therefore a null resultant. The tension is controlled
by inextensibility in a way analogous to pressure in incompressible
hydrodynamics: its discontinuity is a consequence of hydrodynamic
drag, and would be zero in its absence (see
section~\ref{sec:simplemodel}). The torque of the weight, evaluated at
the driving point, is equal to $- \rho_s A_s \ell g \Delta$ where
$\Delta$ is the horizontal distance between the centre of mass of the
string and the driving point. The self sustained state is therefore
characterised by a non vanishing torque due to weight
(figure~\ref{fig:VelocityPressure}d). The effective tension on the other
hand has a null lever arm at the driving point and therefore induces
no torque. The hydrodynamic torque, obtained by integrating the drag
moment along the string, is proportional to the surface area
$\mathcal{S}$ of the loop, and reads
$2 \mathcal{S} \alpha \rho_f R V^2 $. Here we have used Green's
theorem to relate the circulation to the area spanned by the loop:
$| \oint \vec{t}\times \vec{r}\, \diff s | = \oint (x\diff z -
z\diff x) = 2\iint_{\mathcal{S}}\diff x \diff z =
2\,\mathcal{S}$. In the steady state, the hydrodynamic torque
balances the torque due to the weight. It is therefore the
hydrodynamic drag, exerted by the air on the string, which leads to a
self-supporting loop. The horizontal displacement of the centre of
mass $\Delta$ increases with $V$ but must remain below $\ell/4$, by
geometrical constraint. One accordingly predicts that the loop becomes
slim at high speed, its area decreasing as
$\mathcal{S} \propto V^{-2}$.

Under the assumption that further dynamical mechanisms may be safely
ignored, the torque balance can be used to estimate the hydrodynamic
drag from the string shape, namely from the measurement of $\Delta$
and $\mathcal{S}$:
\begin{equation}
 \alpha = \frac{  \rho_s A_s \ell g}{2 \rho_f R V^2 } \frac{\Delta}{\mathcal{S}}
 \label{eq:torquebalance}
\end{equation}
%
%
\begin{figure}
  \centering
  \includegraphics[width=\textwidth]{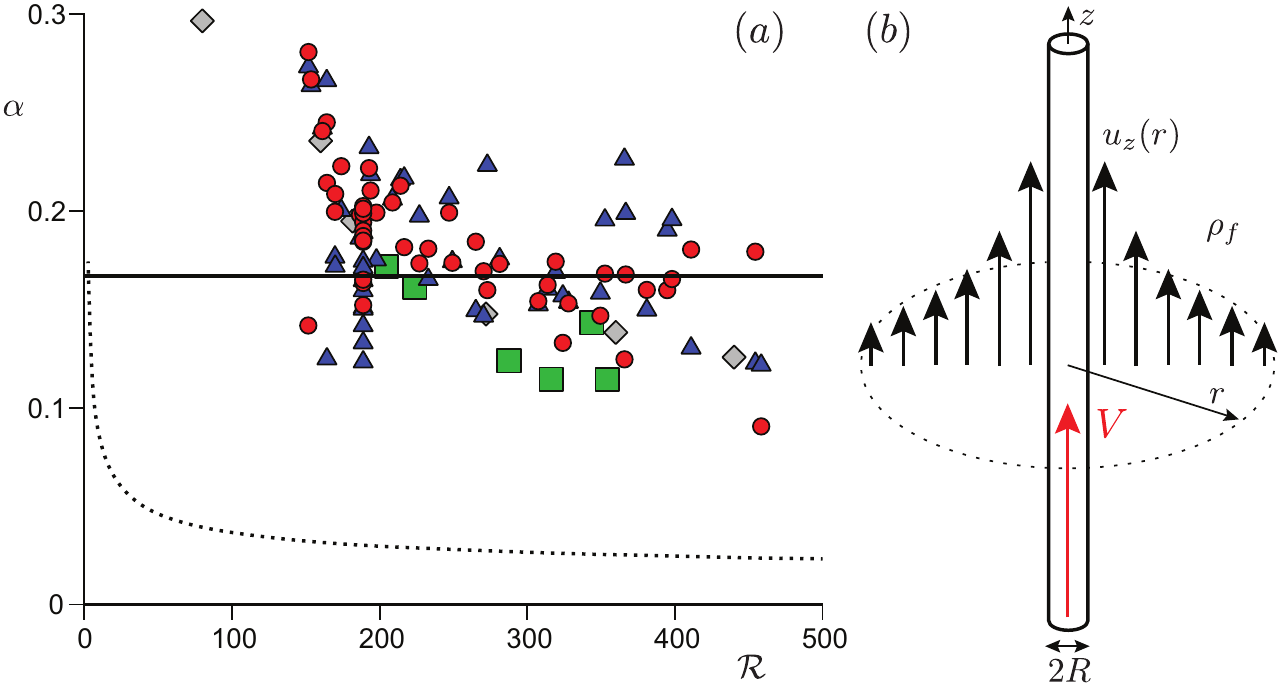}
  \caption{(a) The friction coefficient $\alpha$ decreases slightly as
    a function of the Reynolds number based on the string radius,
    $\mathcal{R}=\rho_fRV/\eta_f$. Red circles:
    equation~\eqref{eq:torquebalance} with $\Delta$ and $\mathcal{S}$ from
    the raw shape, blue triangles: equation~\eqref{eq:torquebalance} with
    $\Delta$ and $\mathcal{S}$ of the numerical best fit of the string
    steady state shape by the theory. The green squares are
    independent measurements on free falling string segments. The grey
    diamonds are measurements on 2-strand acryl yarn reported by
    \citet{uno1972}. The dashed line curve is the parameter-free
    prediction in the hydrodynamically smooth regime (equation~\ref{alpha})
    for an infinite straight string. The solid line is the best fit in
    the hydrodynamically rough regime ($r_0=\SI{180}{\micro\metre}$).
    (b) Notations for the drag calculation. The velocity field is
    assumed to be axisymmetric.}
  \label{fig:AlphaVsRey}
\end{figure}

\subsection{Estimating the turbulent skin drag on the string}
\label{sec:drag}
The effective drag coefficient deduced experimentally from
equation~\eqref{eq:torquebalance} can be compared to theoretical
expectations, based on well established phenomenological laws.
Consider an infinite cylinder moving along its axis at constant
velocity $V$ (figure~\ref{fig:AlphaVsRey}b). This is a turbulent
boundary layer problem in axisymmetric geometry. In the steady state
the momentum flux across any cylinder of radius $r>R$ must be the
same, so that the shear stress $\sigma_{rz}$ decreases as $r^{-1}$.
Using the Prandtl mixing length approach
\citep{schlichting,vanDriest1956,flack2010}, the turbulent Reynolds
stress outside the surface (viscous or rough) sublayer then reads:
\begin{equation}\label{hydrostress}
\sigma_{rz}\simeq \rho_f \kappa^2 (r-R)^2\left|\frac{\partial u_z}{\partial r}\right|\,\frac{\partial u_z}{\partial r}=-\frac{\alpha \rho_f V^2 R}{2\pi r}
\end{equation}
where $\kappa \simeq 0.41$ is the von-K{\'a}rm{\'a}n constant. The
momentum balance integrates into:
\begin{equation}\label{profileu}
u_z=V-\sqrt{\frac{\alpha }{2\pi}} \frac{V}{\kappa} \log\left[\frac{4R}{r_0}\,\frac{\sqrt{r}-\sqrt{R}}{\sqrt{r}+\sqrt{R}}\right]
\end{equation}
where the integration constant $r_0$ is identified as the hydrodynamic
roughness in the limit $r_0\ll R$. In the smooth hydrodynamic
regime, one expects a roughness
$r_0\simeq \frac18 \sqrt{\frac{2\pi}{\alpha}} \frac{\eta_f}{\rho_f V}$
where the factor $1/8$ is a phenomenological constant determined
experimentally in the case of quasi-bidimensional turbulent boundary
layers \citep[chap 17: \emph{logarithmic overlap law}]{schlichting}.
When the viscous sublayer is smaller than the corrugations of the
wire, in the hydrodynamically rough regime, one rather expects $r_0$
to be a fraction of the geometrical roughness. The velocity must
vanish at infinity, which leads to the relation:
\begin{equation}
 \alpha=2\pi \frac{\kappa^2}{\log^2\left[\frac{4R}{r_0}\right]}
 \label{alpha}
\end{equation}
The velocity tends to $0$, far from the string, as $1/\sqrt{r}$. As a
consequence, the momentum per unit length stored in the air,
$\int 2\pi \rho_f r u_z dr$ is infinite in the steady state (the
integral diverges as $r^{3/2}$). This means that added mass effects,
discussed later, would diverge for an asymptotically long and straight
string: they must be limited by the finite size or the finite radius
of curvature, which makes their quantitative modelling difficult.

Figure~\ref{fig:AlphaVsRey} shows that the measured values of
$\alpha$, using equation~\eqref{eq:torquebalance}, are almost
constant, only slightly decreasing with increasing Reynolds number.
The average value $\alpha\simeq 0.167$ corresponds to a hydrodynamic
roughness of $r_0\simeq \SI{180}{\micro\metre}\simeq R/3$, more than
ten times larger than the viscous sublayer. This value is consistent
with rough surface structures whose size is of the order of the string
radius $R$, such as those formed by the three plied strands of
our cotton yarn. The skin friction coefficient $\alpha$ was also
independently measured in a free fall experiment using string segments
attached to an excess weight. This drag is slightly lower, possibly
due to the absence of counter-moving string. Measurements of
longitudinal air-drag of very long cylinders are scarce in the
literature. The air-drag of textile fibres is exploited in air-jet
looms, but studies mostly consider short segments in non-homogeneous
air-flow. As a notable exception \citet{selwood1962}, using an elegant
set-up to drag \SI{20}{\metre} of nylon fibre through still air,
reports a drag coefficient of 0.093 at the highest Reynolds number
$\mathcal{R}=60$, well below our values as expected for a much
smoother filament. Measurements on long ($\ell/R > 4000$) rough
two-filament yarn by \citet{uno1972} on the other hand are in
accordance with our data (figure~\ref{fig:AlphaVsRey}a) and confirm
that plied yarn has a larger effective roughness than mono-filaments.
%
%
\begin{figure}
  \centering
  \includegraphics[width=\textwidth]{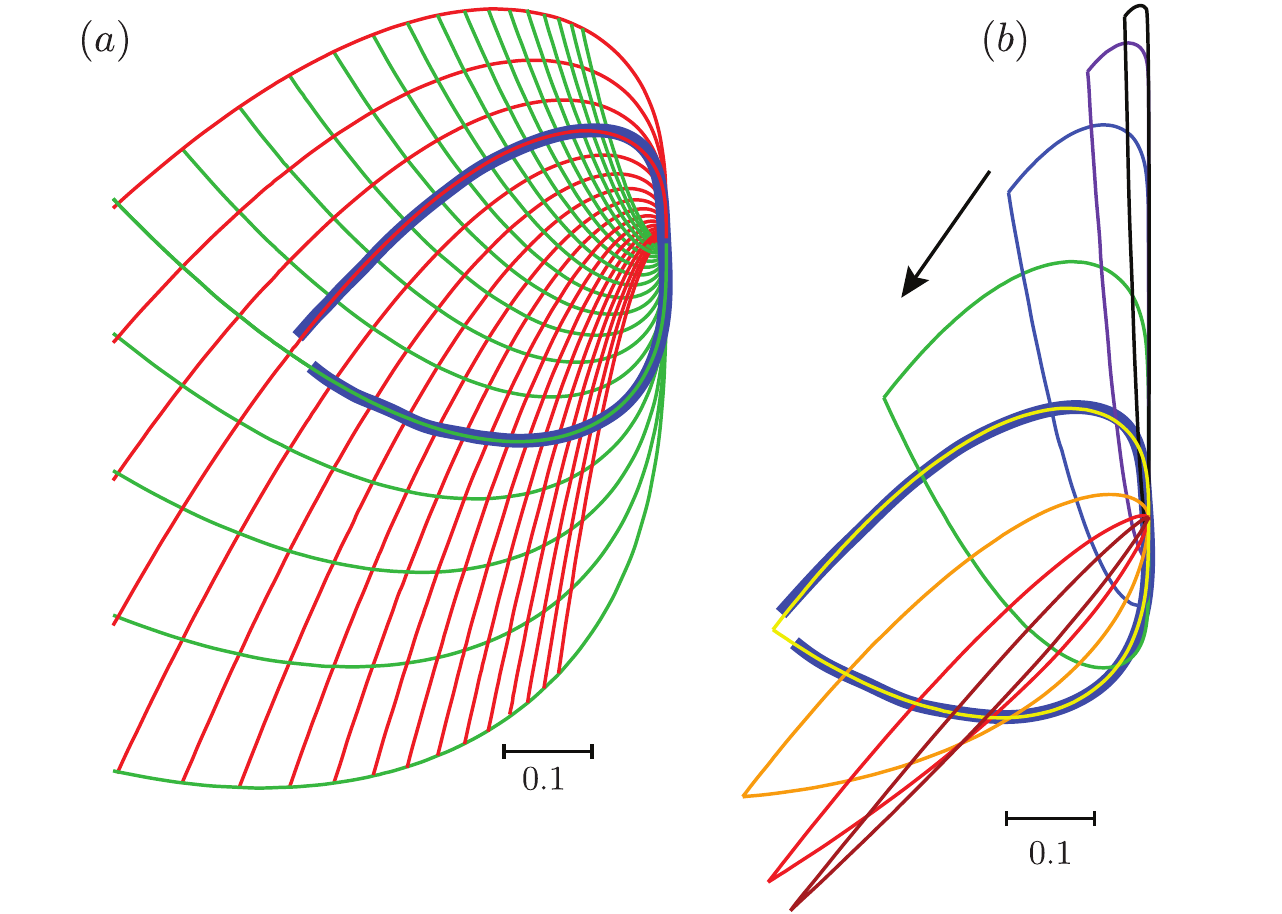}
  \caption{A typical experimental curve observed just above the
    threshold is shown in blue on both panels, obtained for
    $V_0=\SI{5.29}{\metre\per\second}$. (a) Solutions to
    equations~\eqref{simplemodel} for various initial conditions and
    $\mathcal{A} = 1.44$. Red are possible upper branches of the
    steady string loop, green possible lower branches. Note the up
    $\to$ down, red $\to$ green symmetry. In a given experiment (blue)
    the upper and lower parts of the loop each match one of the
    solutions. Note that the propelling system is on the left. Lengths
    are in units of $V^{2}/g$. (b) Change of the selected shape as the
    drag parameter $\mathcal A$ increases (from top to bottom:
    1.00798, 1.02168, 1.05893, 1.1602, 1.43546, 2.18369, 4.21761,
    9.74637), keeping the ejection angle and string length constant.
    The shapes have been superimposed at the singularity, whereas
    experimentally the leftmost point (corresponding to the shooter)
    is fixed. Lengths are in units of $V_0^{2}/g$.}
  \label{fig:Poissons}
\end{figure}

\subsection{Predicted shape}\label{sec:simplemodel}
We express the curvilinear coordinate $s=S V^2/g$ along the string in
terms of its dimensionless counterpart $S$ using $V^2/g$ as the
characteristic length. The equation~\eqref{NS}, projected onto the tangent
and the normal vectors, yields:
\begin{subequations}
 \label{simplemodel}
 \begin{align}
 \label{simplemodel:T}
 \frac{\diff \mathcal{T}}{\diff S} &= \sin \theta + \mathcal{A}\\
 \label{simplemodel:theta}
 \frac{\diff \theta}{\diff S} &= \frac{ \cos \theta}{\mathcal{T}-1}
 \end{align}
\end{subequations}
where the dimensionless drag coefficient scales as $V^2$:
\begin{equation}
\mathcal{A}= \alpha \frac{\rho_f R V^2}{  \rho_s A_s g}
\end{equation}
Equation~\eqref{simplemodel:T} integrates to
$\mathcal{T}\simeq \mathcal{T}_0+\mathcal{A}S+Z$, where $Z$ is the
altitude rescaled by $V^2/g$ and $\mathcal{T}_0$ the tension
immediately after the driving point.

Equation~\eqref{simplemodel:theta} is singular at the critical point
$\mathcal{T}=1$, except if the latter is located at the point whose
tangent is vertical ($\theta=3\pi/2$). In this case,
equations~\eqref{simplemodel} admit around the critical point an asymptotic
solution of the form:
\begin{eqnarray}
\mathcal{T} &\sim& 1+(\mathcal{A}-1)S\\
\theta(s)&\sim& 3\pi/2-C S^{1/(\mathcal{A}-1)}
\end{eqnarray}
This solution is used as a starting point, parameterised by
$\mathcal A$ and $C$, for numerical integration of
equations~\eqref{simplemodel}. Figure~\ref{fig:Poissons}a shows the family
of solutions for a fixed value of the drag $\mathcal A$. Loop shapes
near critical drag (blue example in figure~\ref{fig:Poissons}) are
well reproduced, letting $\mathcal{A}$ be an adjustable parameter.
Figure~\ref{fig:Poissons}b shows solutions that differ in velocity
$V$, but share the same ejection angle and string length. The
succession of shapes shows the transition from gravity dominated
shapes to lifted states.
%
%
\begin{figure}
  \centering
  \includegraphics[width=\textwidth]{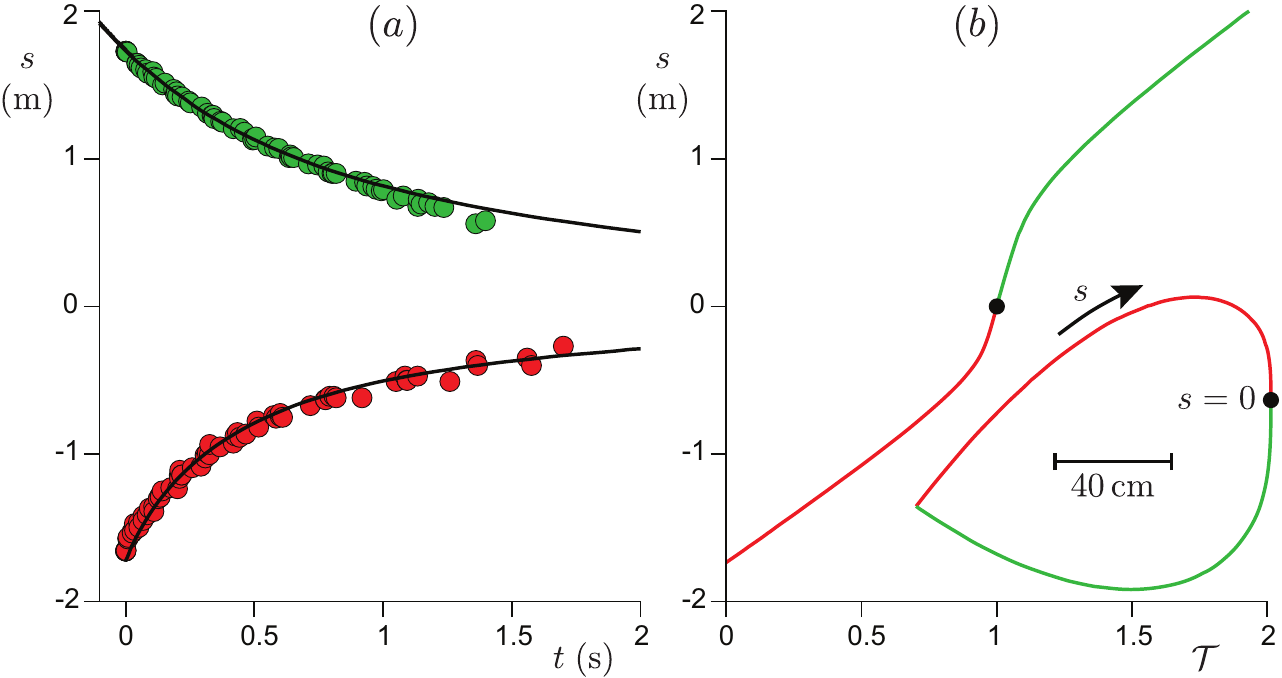}
  \caption{(a) (a) Position of transverse wave-fronts on a string
    above the threshold ($V_0=\SI{5.29}{\metre\per\second}$, same
    experiment as in figure~\ref{fig:Poissons}). Seven deformations were
    tracked on both upper (red) and lower (green) branches, and their
    trajectories superimposed by taking $t=0$ upon leaving the
    propelling wheels. Wave-fronts travelling in the direction of the
    string movement on the upper branch and in the reverse direction
    on the lower branch slow down critically as they approach the
    singularity at $s=0$. The black lines are calculated from the
    tension shown in the right graph. (b) The string tension,
    corresponding to the best shape fit of upper and lower parts by
    our model (inset), increases along the string in the direction of
    string motion. $\mathcal{T}=1$ defines the singular point, which defines the
    origin of the curvilinear coordinate $s$ in both graphs.}
  \label{fig:Ondes}
\end{figure}

\subsection{Wave propagation}
\label{sec:waves}
The theory predicts that all upper branches move at supersonic speeds
with respect to transverse waves ($\mathcal{T}<1$), while on the
contrary lower branches are subsonic ($\mathcal{T}>1$). This produces
a white hole analogue similar to what can be achieved using surface
waves on a flowing fluid \citep{schuetzhold2002,rousseaux2010}. As
opposed to finite time singularities that occur in the motion of
open-ended strings \citep{mcmillen2003,brun2016}, the transsonic
singularity in the propelled loop is stationary. To test its existence
experimentally, we take advantage of the fact that the knot which ties
the string into a loop produces small deformations of the loop shape
at each passage through the propelling wheels. We tracked the position
of seven deformations on both upper and lower branches in one given
experiment, and find that these transverse perturbations exhibit a
critical slowing down as they travel along the string
(figure~\ref{fig:Ondes}a). This is a signature of the singularity:
transverse waves propagate at a speed $\pm \sqrt{T/ \rho_s A_s}$
relative to the moving string. Waves that propagate upstream, that is
in the direction opposite to the string movement, therefore appear to
move at a speed $V - \sqrt{T/\rho_s A_s}$ in the laboratory reference
frame. This speed is positive (moving with the string) on the upper
branch (where $\sqrt{T/\rho_s A_s}=\sqrt{\mathcal{T}}V<V$) and
negative on the lower branch ($\sqrt{T/\rho_s A_s}>V$), so that the
perturbations travel towards the singularity on either side of it. The
best shape fit of this experiment by equations~\eqref{simplemodel}
yields a tension curve $\mathcal{T}(s)$ (figure~\ref{fig:Ondes}b) from
which we deduce trajectories
$\diff s/\diff t = V(1-\sqrt{\mathcal{T}})$ that closely match the
observations (figure~\ref{fig:Ondes}a): this confirms the existence of
the singularity and validates the minimal model, immediately above the
threshold.

\section{Conclusive remarks: beyond the minimal model%
\label{sec:higherordereffects}}
\subsection{Departure from the pear shape}
\begin{figure}
  \centering
  \includegraphics[width=\textwidth]{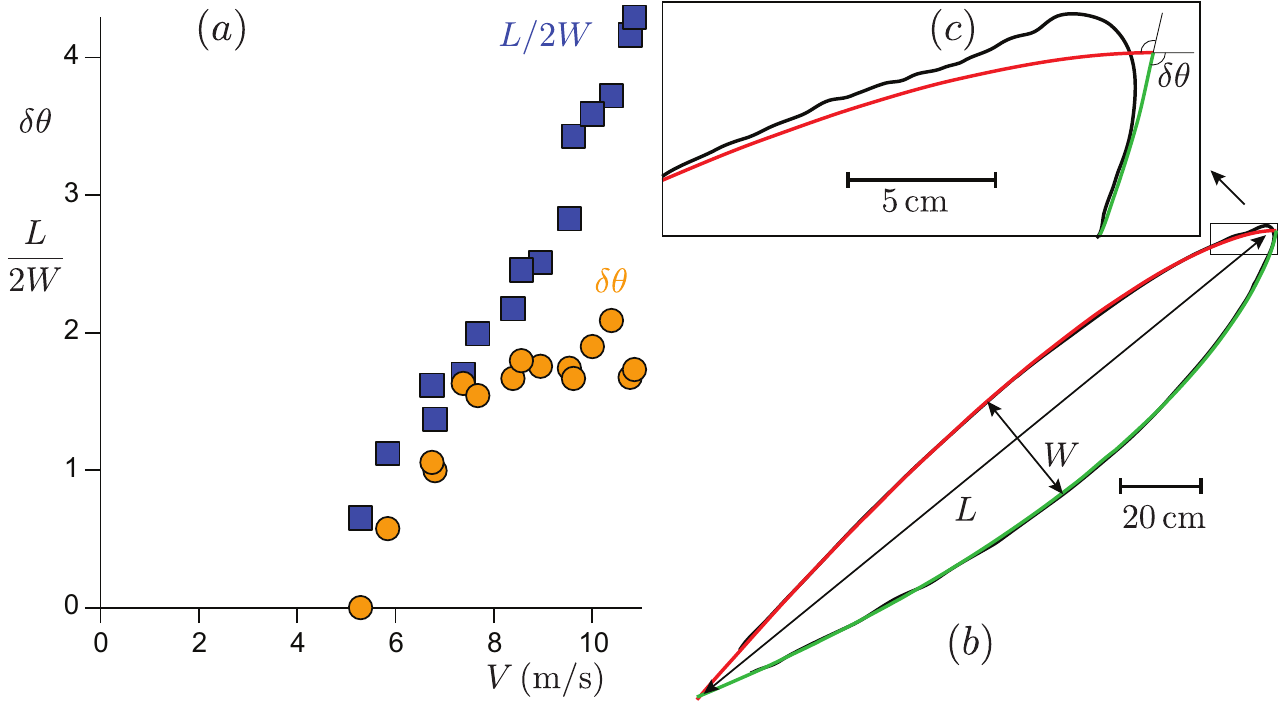}
  \caption{(a) Aspect ratio (blue squares) and angle discontinuity at
    the singular point (yellow circles) of the string loop at
    different velocities. Note the factor $2$ in $L/2W$ added to get a
    common scale. (b) Experimental loop shape (black,
    $\ell = \SI{3.4}{\metre}$, $V=\SI{9.53}{\metre\per\second}$) with
    superposed best fits by the simplified model
    (equations~\ref{simplemodel}) of the upper (red) and lower (green)
    parts. (c) Detailed view of the vicinity of the singularity at the
    top right, showing that the apparent angle discontinuity is
    regularised in the experiment.}
  \label{fig:Discontinuity}
\end{figure}

When the speed is increased above the threshold, the shape becomes
widest in the middle and tapers at both ends in the manner of a
spindle. The spindle becomes slimmer with increasing velocity
(figure~\ref{fig:Discontinuity}a), in accordance with the minimal model
and in particular the torque balance of equation~\eqref{eq:torquebalance}.
At high speed, however, the string exhibits an angular point at the
singularity, which does not correspond to any of the smooth
pear-shaped solutions of figure~\ref{fig:Poissons}. An example is shown
in figure~\ref{fig:Discontinuity}b,c. The upper and lower branches are
still well fitted by equations~\eqref{simplemodel}, on condition that both
the angle and the tension are allowed to be discontinuous across the
singularity. The discontinuity vanishes near the onset of the
self-supporting loop regime (figure~\ref{fig:Discontinuity}a), where the
string has a vertical tangent at the singular point and the shape is
well described by a pear-shaped solution.
%
%

\subsection{Dynamical mechanisms in the vicinity of the critical
  point}
As the upper and lower branches are still well described by the
minimal model outside the class of smooth `pear' solutions, some
further dynamical mechanism must show up and dominate only in the
vicinity of the singularity, and produce the small slightly asymmetric
bend of finite curvature that can be observed in
figure~\ref{fig:Discontinuity}c. Placing an obstacle (a solid plate)
near the tip causes important changes in this curved upper region
around the critical point. This directly shows that non-local effects
are present. Momentum is transferred across the loop via the entrained
fluid, especially between counter-moving upper and lower branches near
the end where the shape tapers. As boundary layers develop along a
straight string, the drag force diminishes \citep{gould1980}. In
contrast we expect the drag coefficient $\alpha$ to increase when the
string is curved, explaining the extra-dissipation associated with the
sudden change of tension around the corner.

\subsection{Added mass effect and hydrodynamic torque}
\label{sec:addedmass}
As air flows along the moving curved string, there must be an added
mass effect i.e. a hydrodynamic force component normal to the string
that can be estimated using a simple momentum balance argument. Let
$\delta$ denote the effective thickness of the air layer that flows at
the speed $V$ of the string and that changes direction when the string
is curved. The momentum flux of this layer is then
$\rho_f \int_{R+r_0}^{R+\delta} 2\pi r u_z^2 \,\diff r \simeq 8\pi
\rho_f R\delta/\log ^2\left(4 R/r_0\right)$, and the centripetal force
required for a direction change is proportional to this flux times
curvature. The reaction on the string can therefore be included by
replacing $\rho_s A_s$ with $\rho_s A_s + \rho_f A_f$, with
$A_f=8 \pi \delta R/\log ^2\left(4 R/r_0\right)$, in the inertial term
of the momentum balance. We will not try here to model $\delta$, which
may depend on the whole shape of the string. As an upper-bound, one
may consider the thickness of a boundary layer developing over a
length $\sim \ell$, obeying $\delta \log (\delta/r_0)\sim \ell$. As
$\delta$ is at most $\sim \SI{10}{\centi\metre}$, the added mass
effect can only increase the apparent density by $\SI{10}{\percent}$
of $\rho_s A_s$.

Additional effects come into play when we consider a string of finite
diameter, for which we may notably have local torque $N$:
\begin{eqnarray}
 \label{withmoments:linearmomentum}
 (\rho_s A_s+\rho_f A_f ) V^2  \frac{\partial }{\partial s}  \vec t &=&\rho_s A_s \vec g + \frac{\partial }{\partial s} (T \vec t + N \vec n) - \alpha \rho_f R V^2 \vec t \\
 \label{withmoments:angularmomentum}
( \rho_s I_s+\rho_f I_f) V^2 \frac{\partial^2 \theta}{\partial s^2} &=&N
\end{eqnarray}
where $I_s=\pi R^4/4$ is the string's moment of inertia and
$I_f=4 \pi \delta^3 R/3\log ^2\left(4 R/r_0\right)$ the added mass
effect, which is a first order description of the hydrodynamic moment
exerted on the string. This time, the added mass effect is typically
$10^3$ larger than the solid contribution (prefactor \num{1e-7} vs
\num{1e-10} in the dimensionless equations), and also larger than the
moment resulting from the string's elastic bending stiffness
\citep{delangre2007}. The latter is estimated from the
elasto-gravitary length $a \simeq \SI{3}{\centi\metre}$ --- measured
from the deflection of cantilevers made of string --- to approximately
$-EI\,g^2/\rho_s A V^6 \simeq -g^3a^3/V^6 \simeq \num{-3e-8}$. Note
that the signs of inertial and elastic stiffness are opposite, so the
relative importance of inertial and elastic stiffness has a strong
impact on the behaviour of the solution near the singularity. The
torque contribution becomes of order 1 when the curvature and its
derivatives change significantly over length scales smaller than
$\num{5e-3}\,V^2/g \simeq \SI{5}{\centi\metre}$, compatible with the
scale at which the string bends near the singularity
(figure~\ref{fig:Discontinuity}c).

These examples of higher order hydrodynamic effects are by no means
exhaustive: the dynamics of finite size bodies accelerating in flows
is notoriously hard to model and includes History forces and drag
corrections \citep{mordant2000,calzavarini2012}. Understanding the
string dynamics in the vicinity of the singularity will thus require a
much more involved analysis of the hydrodynamics of the surrounding
fluid.

\subsection{Perspectives}

The drag on cylinders in axial flow is dominated by the pressure
contributions at the end caps for all but very elongated cylinders.
Loops on the other hand experience only skin drag, making the
experiment discussed here an ideal set-up for its study, including
effects of finite size. A promising application in the textile
industry may be the estimation of axial drag coefficients of fibres
simply from the transition speed between gravitational and lifted
states, and from the string's linear mass. Such a low-tech method may
be of direct interest to textile producers wishing to adjust an
air-jet loom to a new fibre type, without the need for a sophisticated
and costly wind channel and drag measurement apparatus.

On a more fundamental level, the formal resemblance of string dynamics
with 1D hydrodynamics, with tension playing the role of pressure,
suggest a rich field of study both in steady and unsteady cases. The
mathematical structure of the theoretical problem is particularly
interesting, as the dynamical equations are easy to express in
Lagrangian coordinates but the boundary conditions (obstacles for
instance) are typically Eulerian. The creation of lift through axial
drag forces, mediated by tension, is just a hint at more complex
behaviour to be expected in unsteady situations. Of particular
interest may be the interactions between moving filaments in turbulent
flows, and propulsion systems involving several filaments (tentacles).

We thank Physique Exp{\'e}rimentale and the Physics
Department of Universit{\'e} de Paris.

\bibliographystyle{jfm}
\bibliography{JFMStringShooter}

\end{document}